\documentstyle[12pt]{article}

\begin{document}

\input{epsf}

\def\beq{\begin{equation}}
\def\eeq{\end{equation}}
\def\bea{\begin{eqnarray}}
\def\eea{\end{eqnarray}}
\def\beas{\begin{eqnarray*}}
\def\eeas{\end{eqnarray*}}
\def\ov{\overline}
\def\ot{\otimes}

\newcommand{\hf}{\mbox{$\frac{1}{2}$}}
\def\sig{\sigma}
\def\De{\Delta}
\def\af{\alpha}
\def\be{\beta}
\def\la{\lambda}
\def\ga{\gamma}
\def\ep{\epsilon}
\def\vep{\varepsilon}
\def\half{\frac{1}{2}}
\def\third{\frac{1}{3}}
\def\fth{\frac{1}{4}}
\def\sth{\frac{1}{6}}
\def\tth{\frac{1}{24}}
\def\tde{\frac{3}{2}}

\def\zb{{\bar z}} 
\def\psib{{\bar \psi}} 
\def\etab{{\bar \eta }}
\def\gab{{\bar \ga}}
\def\vev#1{\langle #1 \rangle}
\def\inv#1{{1 \over #1}}

\def\CA{{\cal A}}       \def\CB{{\cal B}}       \def\CC{{\cal C}}
\def\CD{{\cal D}}       \def\CE{{\cal E}}       \def\CF{{\cal F}}
\def\CG{{\cal G}}       \def\CH{{\cal H}}       \def\CI{{\cal J}}
\def\CJ{{\cal J}}       \def\CK{{\cal K}}       \def\CL{{\cal L}}
\def\CM{{\cal M}}       \def\CN{{\cal N}}       \def\CO{{\cal O}}
\def\CP{{\cal P}}       \def\CQ{{\cal Q}}       \def\CR{{\cal R}}
\def\CS{{\cal S}}       \def\CT{{\cal T}}       \def\CU{{\cal U}}
\def\CV{{\cal V}}       \def\CW{{\cal W}}       \def\CX{{\cal X}}
\def\CY{{\cal Y}}       \def\CZ{{\cal Z}}

\newcommand{\np}{Nucl. Phys.}
\newcommand{\pl}{Phys. Lett.}
\newcommand{\prl}{Phys. Rev. Lett.}
\newcommand{\cmp}{Commun. Math. Phys.}
\newcommand{\jmp}{J. Math. Phys.}
\newcommand{\jpamg}{J. Phys. {\bf A}: Math. Gen.}
\newcommand{\lmp}{Lett. Math. Phys.}
\newcommand{\ptp}{Prog. Theor. Phys.}

\newif\ifbbB\bbBfalse                
\bbBtrue                             

\ifbbB   
 \message{If you do not have msbm (blackboard bold) fonts,}
 \message{change the option at the top of the text file.}
 \font\blackboard=msbm10 
 \font\blackboards=msbm7 \font\blackboardss=msbm5
 \newfam\black \textfont\black=\blackboard
 \scriptfont\black=\blackboards \scriptscriptfont\black=\blackboardss
 \def\Bbb#1{{\fam\black\relax#1}}
\else
 \def\Bbb{\bf}
\fi

\def\id{{1\! \! 1 }}
\def\bo{{\Bbb 1}}
\def\bI{{\Bbb I}}
\def\bC{{\Bbb C}} 
\def\bZ{{\Bbb Z}}
\def\CN{{\cal N}}

\title{Hubbard Models as Fusion Products of Free Fermions}
\author{{\bf Z. Maassarani}\thanks{Work supported by NSERC 
(Canada) and FCAR (Qu\'ebec).} \\
\\
{\small D\'epartement de Physique, Pav. A-Vachon}\\
{\small Universit\'e Laval,  Ste Foy, Qc,  
G1K 7P4 Canada}\thanks{email address: zmaassar@phy.ulaval.ca} \\}
\date{}
\maketitle

\begin{abstract}
A class of recently introduced $su(n)$ `free-fermion' models has recently 
been  used to construct  
generalized Hubbard models. I derive an  algebra 
defining the `free-fermion' models and give new classes of solutions.   
I  then introduce a conjugation matrix 
and give a new and simple proof of  the corresponding 
decorated Yang-Baxter equation. This provides
the algebraic tools required to couple in an integrable way 
two  copies of free-fermion models. 
Complete integrability of the resulting  Hubbard-like
models is shown by exhibiting  
their  $L$ and $R$  matrices.
Local symmetries of the models are discussed. 
The diagonalization of the free-fermion models is carried 
out using the algebraic Bethe Ansatz.

\end{abstract}
\vspace{5cm}
\noindent
\hspace{1cm} November $11^{\rm th}$ 1997\hfill\\
\hspace*{1cm} LAVAL-PHY-26/97\hfill\\
\hspace*{1cm} cond-mat/9711142

\thispagestyle{empty}

\newpage

\setcounter{page}{1}

\section{Introduction}

The two-dimensional Hubbard model  was introduced to
describe the effects of correlation for $d$-electrons
in transition metals \cite{guhu}. 
It was  then shown to be relevant to the study of 
high-$T_c$ superconductivity of cuprate compounds.

The one-dimensional version also has interesting features.
The model is integrable \cite{liwu,sh12,woa}.
The integrability framework of the model is the quantum inverse scattering 
method \cite{qism1,qism2,qism3}. However, despite sharing 
many properties with
discrete quantum integrable models, the model had a   
peculiar integrable structure which defined  a class of its own.
It was therefore natural to look for integrable generalizations. 
Mapping the fermionic model to a bosonic one with a Jordan-Wigner 
transformation reveals interesting properties. 
The local fermionic symmetries become non-local. This has been known for some time but does not seem to have been  explored further \cite{win}.

Another interesting feature of the 1D Hubbard model and
of most interacting one-dimensional systems is their 
Luttinger liquid behavior \cite{hal,caav}. Such a behavior 
is not restricted to one dimension however \cite{and}.

A bosonic  $su(n)$ Hubbard 
model which contains the usual $su(2)$ model was recently introduced in
\cite{zm1}. These models were shown to be integrable 
and to have an extended $su(n)$ symmetry \cite{zm2}.  
The model is built by coupling two copies of the recently discovered
$su(n)$ XX `free-fermions' models \cite{mm}. For $n=2$ a fermionic
formulation exists, but for $n > 2$ finding an analogous framework 
remains  a tantalizing problem. Strictly speaking, the name 
free-fermions model does
not seem appropriate. It is easy to convince
oneself on dimensional grounds  that an expression in terms of fermionic 
operators can only happen for a subclass of models. 
However for  want of a better characterization
I shall stick to the foregoing appellation.  

Other types of generalizations of the Hubbard model 
exist. They are mostly of the fermionic type, that is, built from fermionic operators; see for example \cite{other1,other2,other3,other4}. 

In this work I look for new solutions of the Yang-Baxter equation
which share the same  features as the known XX models.
I derive  an algebra which unifies the different
`free-fermions' representations, and greatly simplifies the calculations.
This algebra is reminiscent of the Temperley-Lieb algebra.
The former algebra is more restrictive and all the representations
found so far are also representations of the Temperley-Lieb algebra.
Defining a conjugation operator allows for a simple and new derivation of the 
decorated Yang-Baxter
equation. This equation, introduced by Shastry
while studying the usual Hubbard model, is an important algebraic component
of the integrability proof for the `fusion' of two models. 
The `fusion' or coupling of two commuting free-fermions 
copies is then described,
along with an algebraic proof of the integrability of the resulting
Hubbard-like models.  
I then give new representations of the free-fermions  algebra
and of the conjugation matrix. I discuss symmetry issues related to these 
models.  Diagonalization of  the `free-fermions' models
using the algebraic Bethe Ansatz method, shows that their spectrum 
is highly degenerate and simple in a certain sense.
Some outstanding issues and possible directions are discussed in the conclusion. 

\section{A new algebra}\label{algebra}

Let $E^{\af\be}$ be the $n\times n$ matrix with a one at row $\af$ 
and column $\be$ and zeros otherwise. 
Consider  the $\check{R}$-matrix of the $su(n)$ XX model \cite{mm}:
\bea
\check{R}(\lambda) &=&\phantom{+} a(\lambda) \; 
[E^{nn}\otimes E^{nn}+\sum_{{\af, \be<n}}
E^{\af \af}\otimes E^{\be\be}] \nonumber\\
& & + b(\lambda)\;\sum_{{\af<n}}(x E^{\af n}\otimes E^{n\af}   + x^{-1} E^{n\af}\otimes E^{\af n})\nonumber\\
& & + c(\lambda)\; \sum_{{\af<n}}(E^{\af
\af}\otimes E^{nn} + E^{nn}\otimes E^{\af\af}) \label{rcxx}
\eea
where $a(\la)=\cos(\la)$, $b=\sin(\la)$ and $c(\la)=1$.
The functions $a$, $b$ and $c$ satisfy the `free-fermion' condition:
$a^2 +b^2 = c^2$. For this set of parameters,  a Jordan-Wigner
transformation turns the $U=0$ hamiltonian density for $su(2)$ into
a fermionic expression for free fermions hopping on the lattice.
  
I now look for $R$-matrices having the above  form, namely
\beq
\check{R}(\la)= P^{(1)} +P^{(2)} \cos (\la) + P^{(3)} \sin (\la)\label{ansa}
\eeq
and impose the property of regularity
\beq
\check{R}(0)=\bI
\eeq
where $\bI$ is the identity operator. 
One therefore has $P^{(1)}+P^{(2)}=\bI$.
There is no loss of generality in choosing the proportionality 
constant to be one since a solution to the Yang-Baxter equation is 
defined up to a multiplicative function of $\la$.

Requiring $\check{R}$ to satisfy the Yang-Baxter equation
\beq
\check{R}_{12}(\lambda)  \check{R}_{23}(\lambda+\mu)\check{R}_{12}(\mu) =
\check{R}_{23}(\mu) \check{R}_{12}(\lambda+\mu)
\check{R}_{23}(\lambda)\label{ybec}
\eeq
yields a finite set of equations. One develops on a 
set of linearly independent functions and 
equates the operatorial coefficients.  
I do not reproduce here  all the equations
and rather concentrate on the basic ones:
\bea
&P^{(2)}+(P^{(3)})^2=\alpha \;\bI &\label{s1}\\
& [ A+B, [A,B]] = (1-3\alpha )(A-B) +A^3 -B^3&\label{s2}\\
&A^4+(1-2\alpha) A^2= \beta\; \bI\label{s3}
\eea
where $A=P_{12}^{(3)}$, $B=P_{23}^{(3)}$ and $\alpha$, $\beta$ are
two arbitrary complex numbers that arise upon solving equations
of the type $M_{12}=M_{23}$. 
Equation (\ref{s1}) and the regularity equation
allow to keep $P^{(3)}$ as the sole  unknown operator. 
All other equations are therefore equations for $P^{(3)}$. 
Equation (\ref{s2}) means that the Reshetikhin criterion is satisfied 
\cite{resh1,resh2}. This can be seen as an integrability test 
for spin chains \cite{gp1}. 
It seems at first that the large number of constraints on one 
operator cannot have a solution. But we already know that   
(\ref{rcxx}) is a solution for which $\alpha=1$ and $\beta=0$.
I have looked for the minimal set of equations which 
solves all the equations involved in the Yang-Baxter equation and 
have found the following algebra.

Let $E_i\equiv P^{(3)}_{ii+1}$, that is, $E_i$ acts non-trivially on the 
adjacent spaces $i,\, i+1$. The defining relations of the 
free-fermions algebra $\CA$
are:
\bea
& \{ E_i^2, E_{i\pm 1} \} = E_{i\pm 1}\;\;,\;\;\; E_i^3 =E_i &\\
&E_i E_{i\pm 1} E_i =0 \;\;,\;\;\; E_i E_j = E_j E_i \;\;{\rm for}\;\;
|i-j|\geq 2&
\eea
where $\{A,B\}=A B+B A$.
The fourth equation just expresses the fact that 
$E_i$ and $E_j$ commute when they act non-trivially  on disjoint spaces. 
There does not seem to be solutions of the Yang-Baxter
equations for the foregoing $R$-Ansatz unless $\alpha=1$ and $\beta=0$.

The above algebra is reminiscent of the Temperley-Lieb (TL) algebra \cite{tl}.
All the solutions given in section \ref{newmod} can be put, after
a `gauge transformation' (a kind of similarity transformation), 
in a Temperley-Lieb form. Conversely, 
the set of solutions of the Temperley-Lieb algebra is much larger and
most of its solutions do not map to a free-fermions system.
Thus the algebra $\CA$ is much more restrictive.
More details are given in section \ref{newmod}. 

One then checks that $P^{(1)}$ and $P^{(2)}$ 
form a complete set of projectors on the tensor product space 
$\bC^n\otimes\bC^n$:
\beq
P^{(1)} + P^{(2)}=\bI\;,\;\;  (P^{(1)})^2=P^{(1)}\;,\;\; 
(P^{(2)})^2=P^{(2)}\;,\;\; P^{(1)} P^{(2)}=P^{(2)} P^{(1)}=0
\eeq
The operator $P^{(3)}$ is a square root of the operator $P^{(1)}$,
and $P^{(2)}P^{(3)}=P^{(3)}P^{(2)} = 0$.

The above relations imply that the matrix $\check{R}$ 
satisfies the unitarity property
\beq
\check R (\lambda) \check R (-\lambda) = \bI\;\cos^2\lambda
\eeq

\section{The decorated Yang-Baxter equation}\label{conjug}

The decorated Yang-Baxter equation is an equation
similar in form to the Yang-Baxter equation.  It  was first introduced 
in \cite{sh3} as  an algebraic relation at the root of the
integrability of the $su(2)$ bosonic Hubbard model.
This underwent a first generalization in \cite{zm1}.
I give here the ingredients needed for the existence of the DYBE.
 
Assume there exists a `conjugation' matrix $C$ acting on one copy
$\bC^n$ such that
\bea
&C^2=\bI \;,\;\; \{ C_i , P^{(3)}_{12}\}=0 \;\;,\;\;\; i=1,\, 2&\label{rels}\\
&C_1 P^{(3)}_{12} =P^{(3)}_{12} C_2\;\;,\;\;\; 2 (P^{(3)}_{12})^2=\bI-C_1 C_2&
\label{addrel}
\eea
where $C_1=C\otimes\bI$ and $C_2=\bI\otimes C$.
Then $C_i$ commutes with $P^{(1)}_{12}$ and $P^{(2)}_{12}$. 
These relations imply the  following, equivalent, conjugation relations
for $\check{R}$ and $R=\CP \check{R}$, where $\CP$ is the permutation 
operator on $\bC^n\otimes\bC^n$:
\bea
C_i \check{R}_{12}(\la)&=&\check{R}_{12}(-\la) C_i\;\;,\;\;\; i=1,\, 2
\label{crc}\\
C_i R_{12}(\la)&=&R_{12}(-\la) C_j\;\;,\;\;\; i,\, j=1,\, 2\;\;,\;\;\; i\not= j
\eea
Note that only the anticommutator relations were used here.
The remaining relations will be needed in section \ref{trm}.
For the $R$-matrix (\ref{rcxx}) one has 
$C=\sum_{\af < n} E^{\af\af}-E^{nn}$.

That the $R$-matrices satisfy a decorated Yang-Baxter equations (DYBE)
is a simple consequence of the conjugation relations and of the Yang-Baxter
equation.
Consider the following version of the latter equation:
\beq
\check{R}_{12}(\lambda -\mu)  R_{13}(\lambda)R_{23}(\mu) =
R_{13}(\mu) R_{23}(\lambda)\check{R}_{12}(\lambda -\mu)\label{ybec1}
\eeq
One multiplies  (\ref{ybec1}) by $C_1 C_2$, commutes them appropriately
and removes one $C$ using $C^2=\bI$.\footnote{This simple derivation 
of the DYBE does not appear in  \cite{sh3} and appears to have been overlooked
in the literature.}
After letting $\mu\rightarrow -\mu$ one  obtains the 
following  version of the DYBE 
\beq
\check{R}_{12}(\lambda +\mu)\, C_1\, R_{13}(\lambda)\, R_{23}(\mu) =
R_{13}(\mu)\, R_{23}(\lambda)\,C_2\,\check{R}_{12}(\lambda +\mu)\label{dybec}
\eeq
It is worth noting that, while the YBE is invariant under a `gauge
transformation', the DYBE is not. This is due to  the arguments of the matrices.

At this point, it is not clear whether
the existence of $C$ follows from
that of $P^{(3)}$, although it is the case for the models of 
section \ref{newmod}.

In \cite{zm1,zm2} we saw how to couple two $su(n)$ XX models 
\`a la Shastry.  We generalize this procedure to the foregoing models.

\section{A new kind of fusion}

Coupling  two  $su(2)$ XX models in an integrable way in the framework
of the Quantum Inverse Scattering Method was first done by Shastry \cite{sh12}.
It was then generalized in \cite{zm1}.
The results of the above sections provide all the ingredients 
required to couple two solutions of the algebra $\CA$.
Thus the derivation of the two following subsections is algebraic and does 
not depend on a specific representation.

In the context of the quantum inverse scattering method and 
$R$-matrices, the word fusion  has a precise meaning.
It refers to the construction of higher-dimensional solutions
to the Yang-Baxter equation using lower-dimensional solutions.
One uses projection operators and 
reduces the resulting fused-space dimensionality. For instance,
fusing two spin-$\frac{1}{2}$ $su(2)$ $R$-matrices results, after 
projections, in an $R$-matrix for the spin-$1\times$spin-$1$ representation, 
with dimension $3^2\times 3^2$. 

In the kind of fusion described below, one couples two models  
without a  reduction in  dimension for the tensor space of the new model; 
no projection is implemented.
The term fusion here then  takes a different meaning. 
 
\subsection{Lax and transfer matrices}

The transfer matrix is the generating functional of the infinite set 
of conserved quantities.
The construction given in \cite{zm1} is still valid here.
We consider two commuting copies of the free-fermion models found 
in the preceding section. Let us stress that the unprimed and primed
copies need {\it not} be of the same type. For instance, the `left'
copy can be  $(n_1,n_2)$ while the `right' copy is $(n_1^{'},n_2^{'})$
with $n$ not necessarily equal to $n^{'}$ (see section \ref{newmod}
for the definitions and notation).
The Hilbert space of the chain
is a tensor product of $\bC^{n.n^{'}}$, the Hilbert space at each site.

Consider the matrix 
\beq
I_0 (h)  =\cosh (\frac{h}{2})\, \bI + \sinh (\frac{h}{2}) \; C_0 \,C^{'}_0 
=\exp\left(\frac{h}{2}\, C_0 \,C^{'}_0\right)\label{imat}
\eeq
where $C$ and $C^{'}$ are the  conjugation
matrices of the models  being coupled. 
The second equality follows from $C^2=\bI$.  
The parameter
$h$ is related to the spectral parameter $\lambda$ by 
\beq
\sinh (2h) =  U \sin (2\la) \label{rela}
\eeq
where $U$ characterizes the strength of the coupling. 
One chooses for $h(\la)$ the principal branch, which vanishes for 
vanishing $\la$ or $U$.  Then for $U=0$  the monodromy matrix 
(\ref{mono}) becomes a tensor product of the  two, uncoupled,  models.
The Lax operator at site $i$ is given by:
\beq
L_{0i} (\la) = I_0(h)\, R_{0i}(\la)\, R^{'}_{0i}(\la)\, I_0(h)
\eeq
and the monodromy matrix is a product of Lax operators
\beq 
T(\lambda)= L_{0M}(\la)...L_{01}(\la)\label{mono}
\eeq
where $M$ is the number of sites on the chain. 
The transfer matrix is the trace of the monodromy matrix over 
the auxiliary space 0:
\beq
\tau (\la)= {\rm Tr}_0 \;\left[\left( L_{0M}...
L_{01}\right)(\la)\right]\label{tmat}
\eeq
The conserved quantities are then  given by
\beq
H_{p+1} = \left({d^p \ln\tau (\la)\over d\lambda^p}\right)_{\la=0}
\;\;\; , \;\; p\geq 0 \label{cqs}
\eeq
This completes the `fusion' of the two  models. 

A proof that  $H_2$ commutes with  $\tau(\lambda)$ was given in \cite{zm1}.
This proof is algebraic and holds also for the models considered here.
It yields relation (\ref{rela}). The proof in the following section
ensures the complete integrability and yields (\ref{rela}) again. 

The derivative of the matrix $I$ gives the coupling term appearing
in (\ref{h2}). Note that the definition involving a logarithm
has two benefits. Besides giving  the most local operators, it further
disentangles the two copies. 

The construction of a non-additive $R$-matrix intertwining two $L$-matrices 
at different spectral parameters goes through exactly as in \cite{zm2}.
This automatically implies the exact integrability of the models with periodic 
boundary conditions. 

\subsection{The $R$-matrix}\label{trm}

The $R$-matrix which intertwines the Lax operators,
\beq
\check{R}(\la_1,\la_2) \; L(\la_1)\otimes L(\la_2) =  L(\la_2)\otimes L(\la_1)
\;\check{R}(\la_1,\la_2)\label{rll}\;\;,
\eeq 
is again given by 
\bea
\check{R}(\la_1,\la_2) &=& I_{12}(h_2) I_{34}(h_1) \left(
\alpha \,\check{R}_{13}(\la_1-\la_2)  \check{R}_{24}(\la_1-\la_2) 
\right.\nonumber\\
& &\left.+\beta\,  \check{R}_{13}(\la_1+\la_2) C_1 
\check{R}_{24}(\la_1+\la_2)C_2 \right) I_{12}(-h_1) I_{34}(-h_2)
\eea
This matrix acts on the product of four auxiliary spaces labeled from
1 to 4, and $\alpha$, $\beta$ are to be determined.
One then requires relation (\ref{rll}) to be satisfied and 
uses (\ref{ybec1}) and (\ref{dybec})  to derive the following equation:
\beas
&\left(\alpha \,\check{R}_{13}(\la_1-\la_2)  \check{R}_{24}(\la_1-\la_2) 
+\beta\,  C_3\check{R}_{13}(\la_1+\la_2)C_4 \check{R}_{24}(\la_1+\la_2)
\right)I_{12}(2h_1) I_{34}(2h_2) =&\\
&I_{12}(2h_2) I_{34}(2h_1) \left(
\alpha \,\check{R}_{13}(\la_1-\la_2)  \check{R}_{24}(\la_1-\la_2) 
+\beta\,  \check{R}_{13}(\la_1+\la_2) C_1 \check{R}_{24}(\la_1+\la_2)C_2
\right)& 
\eeas
Expanding   the exponentials (see (\ref{imat})) and  
using (\ref{crc}) results in
the cancellation of half the terms on each side.  Using (\ref{ansa})
and relations (\ref{rels}--\ref{addrel}) 
for all the terms, yield only two equations:
\beq
\frac{\beta}{\alpha}= \frac{b}{B}\tanh(h_1+h_2)
\;\;, \;\;\; \;\;
\frac{\beta}{\alpha}= \frac{a}{A}\tanh(h_1-h_2)
\eeq
where $a=\cos(\la_1-\la_2)$, $b=\sin(\la_1-\la_2)$, $A=\cos(\la_1+\la_2)$
and $B=\sin(\la_1+\la_2)$.
The compatibility equation
\beq
\frac{\tan(\la_1-\la_2)}{\tan(\la_1+\la_2)} = \frac{\tanh(h_1-h_2)}{\tanh(h_1+h_2)} 
\eeq 
is satisfied if and only if 
equation (\ref{rela}) is satisfied for the pairs $(\la_1,h_1)$ and
$(\la_2,h_2)$. 
One can then  factor out $\alpha=\alpha(\la_1,\la_2)$ 
which appears as an arbitrary normalization of the $R$-matrix, to obtain:
\bea
\check{R}(\la_1,\la_2)&=& \alpha(\la_1,\la_2) I_{12}(h_2) I_{34}(h_1) \left(
\check{R}_{13}(\la_1-\la_2)  \check{R}_{24}(\la_1-\la_2)
+\frac{\sin(\la_1-\la_2)}{\sin(\la_1+\la_2)}\right.\nonumber\\
& \times&\left. \tanh(h_1+h_2) 
\check{R}_{13}(\la_1+\la_2) C_1 \check{R}_{24}(\la_1+\la_2)C_2 
\right) I_{12}(-h_1) I_{34}(-h_2)\label{rc}
\eea

The underlying algebraic structure at work here is the one elucidated 
in sections \ref{algebra} and \ref{conjug}.
We stress here that this proof is rigorous 
and valid independently of the specific
representation for $P^{(3)}$ and $C$. 
It only  involves  algebraic properties.

We conclude the integrability proof. 
The monodromy matrix being a tensor product 
of $M$ copies of $L$ matrices, one has 
\beq
\check{R}(\la_1,\la_2) \; T(\la_1)\otimes T(\la_2) =  T(\la_2)\otimes T(\la_1)
\;\check{R}(\la_1,\la_2)\label{rtt}
\eeq
Taking the trace over the auxiliary spaces, and using 
the cyclicity property of the trace, one obtains $[\tau(\la_1),\tau(\la_2) ]=0$.
We have thus proven that all the conserved charges $H_p$ mutually commute.

The matrix $\check{R}$ satisfies the regularity property 
\beq
\check{R}(\la_1,\la_1) = \alpha(\la_1,\la_1)\; \bI
\eeq
and the unitarity property:
\bea
\check{R}(\la_1,\la_2) \check{R}(\la_2,\la_1)& =& \alpha^2(\la_1,\la_2)
\cos^2(\la_1-\la_2)\nonumber\\
&\times&\left(\cos^2(\la_1-\la_2) -
\cos^2(\la_1+\la_2) \tanh^2(h_1-h_2)\right) \bI
\eea
The intertwiner $\check{R}$  satisfies a Yang-Baxter relation 
of its own:
\beq
\check{R}_{12}(\la_2,\la_3) \check{R}_{23}(\la_1,\la_3)
\check{R}_{12}(\la_1,\la_2) = \check{R}_{23}(\la_1,\la_2) 
\check{R}_{12}(\la_1,\la_3) \check{R}_{23}(\la_2,\la_3)\label{rybe}
\eeq
where $\la$ and $h$ are related through (\ref{rela}).
As explained in \cite{zm2} the direct verification of this relation 
is tedious,
but can be avoided. The proof done for $su(2)$ in \cite{sw} generalizes;
this proof is based on Korepanov's tetrahedral Zamolodchikov 
algebra \cite{korep}.  
All variants of the  proof follow from the 
foregoing algebraic structure. 
A notable feature of the matrix $\check{R}(\la_1,\la_2)$
is its non-additivity property; the 
$\la$ dependence cannot be reduced to a difference $(\la_1-\la_2)$.
This is the source of the difficulty in verifying (\ref{rybe}). 

I now give new solutions of the YBE which have the foregoing properties.

\section{New models}\label{newmod}

Let $n$, $n_1$ and $n_2$ be three integers such that
\beq
2\leq n \;\;,\;\;\; 1\leq n_1\leq n_2\leq n-1 \;\;,\;\;\; n_1+n_2=n
\eeq
and $A$, $B$ be two disjoint sets whose union is the set of basis 
states of $\bC^n$, with card$(A)=n_1$ and card$(B)=n_2$. 
Let 
\bea
P^{(3)}&=&\sum_{a\in A}\sum_{\beta\in B}\left(x_{a\beta} E^{\beta a}\otimes
E^{a\beta} + x_{a\beta}^{-1} E^{a\beta}\otimes E^{\beta a}\right)\\
P^{(1)}&=&(P^{(3)})^2=\sum_{a\in A}\sum_{\beta\in B}\left(E^{\beta\beta}\otimes
E^{a a} + E^{a a}\otimes E^{\beta\beta}\right)\\
P^{(2)}&=&\bI - P^{(1)}=\sum_{a,a^{'}\in A} E^{a a}\otimes
E^{a^{'} a^{'}} + \sum_{\beta,\beta^{'}
\in B} E^{\beta\beta}\otimes E^{\beta^{'} \beta^{'}}
\eea
The $n_1 . n_2$ parameters $x_{a\beta}$ are arbitrary complex numbers. 
Latin indices belong to $A$ while greek indices belong to $B$.
These operators satisfy all the defining relations of the algebra $\CA$
and therefore one has an $R$-matrix
\beq
\check{R}(\la)= P^{(1)} +P^{(2)} \cos (\la) + P^{(3)} \sin (\la)
\eeq
which satisfies the YBE. Denote this representation by $(n_1,n_2)$.

The conjugation matrix is defined up to an overall sign; it is given by
\beq
C=\sum_{\beta\in B} E^{\beta\beta} -\sum_{a\in A} E^{aa}\label{cc}
\eeq
and satisfies all the  relations of the preceding section. 
The DYBE therefore holds. 

Note that, unless one wants to perform numerical calculations 
or write down an $n^2\times n^2$ matrix representation, one need 
not specify which states belong to which set, thereby keeping
a `symmetrical' labeling. Note also that the restriction 
$n_1\leq n_2$ is not essential. It just avoids a double counting
of distinct models since one has the obvious symmetry $A\leftrightarrow B$.
The number of models, for a given $n$, is equal to the integer
part of $n/2$: $[\frac{n}{2}]$. 

It is possible to perform a gauge transformation which puts
these models in a TL form.\footnote{The connection between some XX models
and their TL formulation was pointed out by  Martins \cite{mjmp}.} 
One obtains 
\bea
\check{R}^{\rm TL}(\la)&=&\bI \,\cos (\la) + (P^{(3)} + P^{\rm TL})\,\sin(\la)\\
P^{\rm TL}&=&i\sum_{{a\in A}}\sum_{{\be\in B}}
\left( E^{\be\be}\otimes E^{aa}- 
E^{aa}\otimes E^{\be\be} \right)
\eea
where $i^2=-1$. 
The operator $E=P^{(3)} + P^{\rm TL}$ satisfies the Temperley-Lieb
algebra
\beq
E_i^2=0\;,\;\; E_i E_{i\pm 1}E_i =E_i\;,\;\; E_i E_j=E_j E_i \;\;{\rm for}
\;\;\; |i-j| \geq 2
\eeq
and $\check{R}^{{\rm TL}}(\la)$ the Yang-Baxter equation.
Let $M$ be any invertible matrix. A general class of solution 
of the TL algebra is given by
\beq
E_{ab,cd}= M_{ab} M^{-1}_{cd}\;\;, \;\; {\rm tr} ({}^t\! M.M^{-1})=0
\eeq
where, on the left-hand side,  $a,b$ ($c,d$) are the row (column) indices. 
However, only for $n=2$ and $M\propto \,{\rm antidiag}(1,\pm i)$ 
does one have a free-fermions model.
The other $(n_1,n_2)$ models do not fit in this scheme. 
This confirms the statement made in section \ref{algebra}. 
 
For $n_1=1$, $n_2=n-1$, and all the $x_{a\beta}$ equal to each other,
one recovers the  $su(n)$ XX models
found in \cite{mm}. Allowing the twist parameters $x$ to be unequal amounts 
to a multiple deformation of these models. We see in the next
section that the degree of symmetry depends on the $x$'s.

Before diagonalizing the conserved quantities of the free-fermion 
models we pause to consider their symmetries and 
their  quadratic defining hamiltonians.

\section{Hamiltonians and symmetries}
 
Periodic boundary conditions are assumed.  
Consider first the free-fermions models.
The quadratic hamiltonian calculated by (\ref{cqs}) is given by
\beq
H_2= \sum_i P^{(3)}_{ii+1}
\eeq 
The cubic hamiltonian is equal to
\beq
H_3 = \sum_i \, [ P^{(3)}_{ii+1}, P^{(3)}_{i+1i+2} ]
\eeq
This is a boosted form of $H_2$ \cite{boost}. The commutation of $H_2$
and $H_3$ is a consequence of the Reshetikhin
criterion (\ref{s2}). We expect considerations
about the boost structure and the explicit form 
of the conserved quantities 
to generalize straightforwardly \cite{mm}.

The quadratic Hubbard-like hamiltonians obtained 
by fusion  are given by
\beq
H_2=\sum_i P^{(3)}_{ii+1} +\sum_i P^{'(3)}_{ii+1} + 
U\sum_i C_i C_i^{'}\label{h2}
\eeq
where  primed and unprimed quantities correspond to the two  commuting   
copies. The cubic hamiltonian is not given by a boosted
form of $H_2$ \cite{zm1,gp2}.
 
The hamiltonians $H_2$ are defined
in one dimension but can be evidently defined on any lattice; integrability is
lost however. 
These  hamiltonians   can be written simply  
in terms of  $su(n)$ hermitian traceless matrices.
For $|x_{a\beta}|=1$ and $U$ real the hamiltonians are hermitian.

Because of the structure of the hamiltonians one expects, at least, 
the diagonal generators to commute with all the conserved quantities.
As seen for the models of \cite{mm} the symmetry may be larger.
One has the following relations
\bea
\forall a,b \in A\;\; \;{[ E_i^{ab}, L_{0i}(\la) ]} 
&=& -{[ E_0^{ab}, L_{0i}(\la) ]}
\;\;\; {\rm iff}\;\; x_{a\be}=x_{b\be} \;\;\forall 
\be \in B\label{cm1}\\
\forall \af,\be \in B\;\;\;\;{[ E_i^{\af\be}, L_{0i}(\la) ]} 
&=& -{[ E_0^{\af\be}, L_{0i}(\la) ]}
\;\;\; {\rm iff}\;\; x_{a\af}=x_{a\be} \;\;\forall 
a \in A\label{cm2}
\eea
where $L=R$.
The linear magnetic-field operators
\bea
H_1^{ab}&=& \sum_i E_i^{ab}\;\;\; a,b \in A\\
H_1^{\af\be}&=& \sum_i E_i^{\af\be}\;\;\; \af,\be  \in B
\eea
commute with the transfer matrix if the parameters $x_{a\be}$ satisfy the
above constraints for the corresponding indices. One just uses
the expression (\ref{tmat}),  its cyclic structure and the
relations (\ref{cm1}) and (\ref{cm2}).
The commutation with $\tau(\la)$ implies the commutation
with all the hamiltonians $H_p$. 

In particular all the diagonal operators are symmetries, without 
any constraints on the parameters $x$. 
When all the parameters are equal to one parameter, say $x_{a\be}=x\,, 
\;\forall a\in A\;,\; \forall\be\in B$,
the full local symmetry is $su(n_1)\oplus su(n_2)\oplus u(1)$.
This symmetry is largest (in terms of number of generators)
for $n_1=1$ and $n_2=n-1$, that is for the models of  reference \cite{mm}.

It is straightforward to conclude that the models  obtained by fusion
inherit the local symmetries of their components. Again,
when  $x_{a\af}^{(l)}=x^{(l)}$ for the left copy and $x_{b\be}^{(r)}=x^{(r)}$
for the right copy, the full local symmetry of the 
model $(n_1,n_2) \times
(n_1^{'},n_2^{'})$ is 
$\left( su(n_1)\oplus su(n_2)\oplus u(1)\right)\times
\left( su(n_1^{'})\oplus su(n_2^{'})\oplus u(1)\right)$.

One consequence is that one can add   magnetic-field terms
for each symmetry generators, without
spoiling the integrability of the models.

\section{Algebraic Bethe Ansatz} 

The diagonalization by Bethe Ansatz of the free-fermions models
is very similar to the one for the $su(n)$  XX models. See
\cite{mm} for additional details omitted  here.
I am considering the case $x_{a\be}=x$ to avoid unessential complications.

The transfer matrix  defined in (\ref{tmat}) is the trace over the auxiliary
space  of the monodromy matrix $T(\lambda)$. The latter is an $n$-dimensional 
matrix whose entries are operators acting on the Hilbert space $C^n\otimes...
\otimes C^n$, with a copy for every site. The number of sites
is $M$. Some elements of the monodromy
matrix are used to create Ans\"atze for the eigenstates. 
When written in components, equation (\ref{rtt})  provides 
the algebraic relations 
needed to find the action 
of the transfer matrix on the states.

We now use the following notation for some elements
of the monodromy matrix: 
\beq
S=T_{11}\,, \; C_a=T_{1a} \;,\;\; a=2,...,n_1 \;, \;\; 
C_\be=T_{1\be} \;,\;\; \be=n_1+1,...,n 
\eeq
where, as usual, the Latin indices belong to $A$ and the greek indices to $B$.
The remaining elements are denoted by $T_{**}$.
The transfer matrix is given by 
\beq
\tau(\lambda)= S(\lambda) +\sum_{a=2}^{n_1} 
T_{aa}(\lambda) +\sum_{\be=n_1+1}^n T_{\be\be}(\la)
\eeq

It is easy to see that the
vector $||1\rangle \equiv |1\rangle\otimes ... \otimes|1\rangle$ 
is an eigenvector of the transfer matrix.
The only non-vanishing elements of the monodromy 
matrix on $||1\rangle$ are:
\beq
S(\la)\,||1\rangle= (\cos(\la))^M||1\rangle\;,\;\; 
T_{\be\be}(\la)\,||1\rangle=(x^{-1}\sin(\la))^M ||1\rangle
\eeq
and the action of {\it all} the $C$ operators, $C_a$ ($a\not= 1$) and
$C_\be$.

It turns out that it is still  possible to construct Bethe Ansatz eigenvectors
using the $C_{\be}$ only, namely:
\beq
|\lambda_1,..., \lambda_p\rangle =
\sum_{\af_p,..., \af_1} F^{\af_p,..., \af_1}
C_{\af_1}(\lambda_1)...C_{\af_p}(\lambda_p) \; ||1\rangle
\eeq 
where the parameters
$\lambda_i$ and the coefficients $F$ are to be determined.
This Ansatz also vanishes identically if $p>M$; the proof of this fact
is given below.
Equation (\ref{rtt}) gives the following relations:
\bea
& C_\af(\lambda) C_\be(\mu) =C_\rho(\mu) C_\sigma(\lambda) 
\CP_{\sigma\rho,\af\be}&\\
& S(\la)C_\be(\mu) = f(\mu-\la) C_\be(\mu) S(\la) + 
g(\mu-\la) C_\be(\la) S(\mu)&\\
& t_{\af\be}(\la)C_\ga(\mu) = f(\la-\mu)C_\rho(\mu) t_{\af\sigma}(\la) 
\CP_{\sigma\rho,\be\ga} + 
g(\la-\mu) C_\be(\la) t_{\af\ga}(\mu)&\\ 
& f(\la)= {x\cos \la \over  \sin \la}\;, \;\;\; 
g(\la)=-{x\over \sin \la}\;,&
\eea 
where  $\CP$ is the permutation
operator for $\bC^{n_2}\otimes\bC^{n_2}$. 
It is important to notice that the Latin and greek indices 
do not mix in these relations.

One then applies the transfer matrix on the state $|\lambda_1,...,
\lambda_p\rangle$ and with the help of the above relations commutes 
it through the $C_\af$'s. The contributions from $S$ and $T_{\be\be}$
are treated exactly as in \cite{mm}. The contributions
from $T_{aa}$ ($a\not=1$) vanish for $p<M$ while 
$|\lambda_1,..., \lambda_p\rangle$ is an eigenvector when $p=M$,
without any constraint. 
To see this we need the commutation relations between the 
Cartan subalgebra generators and the creation operators $C$.
Using relations (\ref{mono}), (\ref{cm1}) and (\ref{cm2}), 
one easily derives
\bea
 {[ H_1^{11}, C_\be (\la) ]}&=& -  C_\be (\la)\label{huuc} \\
 {[ H_1^{aa}, C_\be (\la) ]}&=& 0\;,\;\;\; \forall a\in A-\{ 1\}
\label{haac}
\eea
Relations (\ref{haac}) imply that the eigenvector Ansatz
has no $a$-states in it ($a\not= 1$), while relation (\ref{huuc}) implies
that $C_{\af_1}(\lambda_1)...C_{\af_p}(\lambda_p) \; ||1\rangle$ has 
only $B$-states in it when $p=M$.
This means
\bea
C_{\af_1}(\lambda_1)...C_{\af_p}(\lambda_p) \; ||1\rangle &\equiv& 0
\;\; {\rm for} \;\; p>M\\
T_{aa}(\la)\,  C_{\af_1}(\lambda_1)...C_{\af_p}
(\lambda_p) \; ||1\rangle &=& 0\;,\;\;
\; 0 \leq p\leq M-1\;\; {\rm and}\;\; a\not= 1\\
T_{aa}(\la)\,  C_{\af_1}(\lambda_1)...C_{\af_M}
(\lambda_M) \; ||1\rangle &=& (x \sin(\la))^M
C_{\af_1}(\lambda_1)...C_{\af_M}(\lambda_M)||1\rangle\;,\;\;
\; a\not= 1
\eea

One then finds the corresponding 
eigenvalues of $\tau(\la)$
\bea
\Lambda^{((n_1,n_2),M)}(\la)&=&(\cos(\la))^M\prod_{j=1}^p f(\la_j-\la) + 
(x^{-1}\sin(\la))^M \left(\prod_{j=1}^p f(\la-\la_j)\right)
\Lambda^{(n_2,p)} \nonumber\\
& \phantom{+}& +  (n_1-1) (x\sin(\la))^M \delta_{pM}\label{lamb}
\eea
Here $\Lambda^{(n_2,p)}$ is an eigenvalue of the 
unit-shift operator $\tau^{(n_2,p)}$, 
for a chain of $p$ sites and $n_2$ possible states at each site; 
it is constructed out of  the
permutation operator $\CP$ on $\bC^{n_2}\otimes\bC^{n_2}$.
The coefficients $F^{\af_p,..., \af_1}$ are such $F$ is an eigenvector
of $\tau^{(n_2,p)}$ for the eigenvalue $\Lambda^{(n_2,p)}$.
Finally the Bethe Ansatz equations are just
\beq
(-1)^{p-1} \left({ x \cos(\la_j)\over \sin(\la_j)}\right)^M=\Lambda^{(n_2,p)}
\;\;\;\; , \; j=1,...,p
\eeq
The operator $\tau^{(n_2,p)}$ can be written as a product 
of disjoint  permutation cycles. One also  
has $\left(\tau^{(n_2,p)}\right)^p= \bI$ .  
The eigenvalues $\Lambda^{(n_2,p)}$ 
are then, {\it at most},
$p^{\rm th}$ roots of unity and are highly degenerate. The dimensions of the 
cycles and their multiplicities
will depend on both $p$ and $n_2$. 

One can perform the above diagonalization procedure over the pseudo-vacuum
$||a\rangle$ ($a\not=1$). The set of eigenvalues is exactly the same
as the one found above and the eigenvectors have the same structure
but form a completely  distinct set, at least for $0\leq p < M$. 
This is easily inferred from the action of the Cartan generators. 
One can also start with any of the vectors $||\be\rangle$, $\be\in B$
and obtain yet other sets of eigenvectors. 
The superscript $n_2$ is replaced by $n_1$ in (\ref{lamb}), 
$\Lambda^{(n_2,p)}$  and  $\tau^{(n_2,p)}$.
These features reflect a large degeneracy of the spectrum. 

Finally, for $n>2$,   the 
eigenvectors $|\lambda_1,..., \lambda_p\rangle$
are generically not eigenvectors of all the magnetic field
operators $H_1^{**}$ \cite{mm}. Because the spectrum is degenerate 
this is not in contradiction with the fact that the magnetic 
operators commute with the conserved quantities.

We have thus  diagonalized the transfer matrix. 
The nested Bethe Ansatz truncates and the spectrum is trivial in the 
above sense.

\section{Conclusion}

A new algebra defining bosonic integrable `free-fermions' representations
has been derived.   These models were shown to have a highly degenerate
and `simple' spectrum and to possess local symmetries. Another 
distinguishing feature of these models is the possibility to couple 
any two of them in an {\it integrable} way. The algebraic structure at
the root of this `fusion' has been put in a simple algebraic 
setting, thereby unifying and simplifying all the derivations. 
New representations were found, generalizing the known
XX models \cite{mm} and the Hubbard-like models \cite{zm1}.
 
Finding a multiparametric deformation
of $su(n)$ interpolating between the XXZ models and the models considered
here would be interesting \cite{ar,mm}. This would provide
a quantum group structure which would shed a new light on the
models at hand.

Integrable bosonic Bariev chains and their multicopy generalizations 
have been found \cite{bar,zhou}. These chains, in
their fermionic formulation describe correlated hoppings of
electrons on a chain.  The bosonic chains are 
obtained by coupling, in a  yet different way, two or more
$su(2)$ XX models. It is therefore likely that one can 
use the free-fermions models to build new, more general Bariev chains.
Martins reported some progress in this direction \cite{mjmp}.
The algebra found here is bound to describe the 
algebraic structure at work for the Bariev chains.

\bigskip\ {\bf Acknowledgement:} I thank M.J. Martins for interesting
and motivating discussions, 
M. Shiroishi for communications and for bringing to my attention
reference \cite{sw}. I am grateful for the continued support of  P. Mathieu.

\end{document}